\documentclass[journal]{IEEEtran}
\usepackage{multirow}
\usepackage{mathrsfs}
\usepackage{amsfonts}
\usepackage{amssymb}
\usepackage{graphicx}
\ifCLASSINFOpdf
\else
\fi
\usepackage[cmex10]{amsmath}
\usepackage{amssymb}  % assumes amsmath package installed
\usepackage{amsfonts}

\usepackage{mdwmath}
\usepackage{mdwtab}
\begin{document}
%
% paper title
% can use linebreaks \\ within to get better formatting as desired
\title{Optimal identification of non-Markovian environments for spin chains}

\author{Shibei Xue~\IEEEmembership{Member,~IEEE,} Jun Zhang~\IEEEmembership{Senior Member,~IEEE,} and~Ian~R.~Petersen~\IEEEmembership{Fellow,~IEEE}
%\thanks{This research was supported under Australian Research Council��s Discovery
%Projects and Laureate Fellowships funding schemes (Projects DP140101779 and FL110100020) and the Chinese Academy of Sciences President's International Fellowship Initiative (No. 2015DT006).}
\thanks{This research was supported by the National Natural Science Foundation of China (No. 61873162) and Shanghai Pujiang Program (Grant No. 18PJ1405500). This research was also supported under Australian Research
Council Discovery Projects and Laureate Fellowships funding
schemes (Projects DP140101779, DP180101805, and FL110100020) and
the Air Force Office of Scientific Research (AFOSR) under
agreement FA2386-16-1-4065.
%MJ thanks the National Natural
%Science Foundation of China (No. 61473199 and 61104002).
JZ thanks the National Natural Science Foundation of China
(No. 61673264 and 61533012), and State Key Laboratory of
Precision Spectroscopy, ECNU, China.}
\thanks{S. Xue is with Department of Automation, Shanghai Jiao Tong University and Key Laboratory of System Control and Information Processing, Ministry of Education of China, Shanghai 200240, China. (e-mail: shbxue@sjtu.edu.cn)}
\thanks{J. Zhang is with the Joint Institute of UM-SJTU and Key Laboratory of System Control and Information Processing, Ministry of Education of China, Shanghai 200240, China. (e-mail: zhangjun12@sjtu.edu.cn)}
\thanks{I. R. Petersen is with Research School of Engineering, Australian National University, Canberra, ACT 2600, Australia.(e-mail: i.r.petersen@gmail.com)}}% i.r.petersen@gmail.com).

\maketitle

\begin{abstract}
%\boldmath
Correlations of an environment are crucial for the dynamics of non-Markovian quantum systems, which may not be known in advance. In this paper, we propose a gradient algorithm for identifying the correlations in terms of  time-varying damping rate functions in a time-convolution-less master equation for spin chains. By measuring time trace observables of the system, the identification procedure can be formulated as an optimization problem. The gradient algorithm is designed based on a calculation of the derivative of an objective function with respect to the damping rate functions, whose effectiveness is shown in a comparison to a differential approach for a two-qubit spin chain.
%\textbf{with which power spectral density of the environment can be identified.
\end{abstract}
% IEEEtran.cls defaults to using nonbold math in the Abstract.
% This preserves the distinction between vectors and scalars. However,
% if the conference you are submitting to favors bold math in the abstract,
% then you can use LaTeX's standard command \boldmath at the very start
% of the abstract to achieve this. Many IEEE journals/conferences frown on
% math in the abstract anyway.

% no keywords

% For peer review papers, you can put extra information on the cover
% page as needed:
% \ifCLASSOPTIONpeerreview
% \begin{center} \bfseries EDICS Category: 3-BBND \end{center}
% \fi
%
% For peerreview papers, this IEEEtran command inserts a page break and
% creates the second title. It will be ignored for other modes.
\IEEEpeerreviewmaketitle

\section{Introduction}
To exactly process quantum information, accurate models, including parameters, structures, and descriptions of dynamics, for quantum information carriers are required. With these models, sophisticated feedback control strategies can be designed, for example, feedback stabilization of a number state in a cavity~\cite{HarocheNAT2011}, preservation of quantum coherence and entanglement for qubit systems~\cite{5395610} and linear quantum systems~\cite{XueCST2016}, or coherent feedback rejection of quantum colored noise~\cite{XuePRA2012,xuefermion}.

However, in practice, an accurate model may not be obtainable since parts of the parameters, structures or dynamics of the quantum system may not be well understood. This would lead to unexpected experimental results or degraded control performance of a quantum control system. For example, in a recent experiment on quantum dots~\cite{Frey}, a calculation based on the theoretical model has a discrepancy from the experimental data on the broadened resonator line-width under suitable parameters, which means that some unknown dynamics of the system are not included in the model. Correspondingly, it is shown that degraded estimation performance can be observed due to ignorance of a noise model for a quantum system~\cite{xueACC2016}. Hence, the problem of how to
determine the unknown parameters, structures or dynamics of a ``dark" quantum system is an essential problem in quantum control theory which is referred to quantum system identification.

A general identification methodology involves finding the unknown parts of a quantum system from measurement data (e.g., the spectrum of an output field or expectations of observables,) extracted from the system under excitation~\cite{PhysRevA.87.022324}. The identification method was firstly explored to extract Hamiltonian information for closed quantum systems~\cite{PhysRevLett.89.263902, PhysRevLett.108.080502, JunPRL14}. Moreover, it is worth considering the identification problem for open quantum systems where the quantum system is disturbed by quantum noises arising from an environment. When the correlation function of the quantum noise is the Dirac delta function; i.e., the correlation time of the quantum noise is very short, the noise and the relevant quantum system refer to quantum white noise and a Markovian quantum system, respectively~\cite{Breuer}. For Markovian quantum systems, a continuous-measurement-based method was proposed to identify unknown parameters in a cavity-atomic system~\cite{Mbu96QSO} and  unknown structures of a spin network~\cite{NaokiNJP14}. Also, a system-realization-theory-based method in~\cite{JunPRL14} was extended to  estimate unknown parameters in the Hamiltonian of a Markovian spin network from measurement time traces~\cite{JunPRA15}, whose identifiability was discussed in~\cite{sonePRA17}.
%\cite{PhysRevA.81.063422},
In addition, an identification method for linear Markovian quantum systems was systematically discussed in~\cite{NaokiTAC16}.

In contrast to the Dirac correlation function of quantum white noise, correlation functions of quantum colored noise can be more complicated due to the memory effect of non-Markovian environments. A quantum system disturbed by quantum colored noise is referred to as a non-Markovian quantum system, whose dynamics is quite different from that of the Markovian quantum systems~\cite{Breuer}. To control non-Markovian quantum systems, it is crucial to acquire the knowledge of the correlation functions of the non-Markovian environment beforehand.
 Wu {\it et al.} proposed a frequency domain approach to the identification of the environment spectrum for a superconducting single qubit at an optimal point~\cite{PhysRevA.87.022324}. For the quantum dot system in~\cite{Frey}, an augmented system model was presented to explore the structure of the non-Markovian environment~\cite{Xue17ac}. Also, a spectroscopic method was presented to explore the spectrum of spin baths both theoretically~\cite{PhysRevA.95.022121} and experimentally~\cite{PhysRevLett.118.177702}, which involves high-energy dynamical decoupling pulses.

In this paper, we aim to extract the correlation functions of non-Markovian environments for spin chains.
The dynamics of the spin chains are described by a time-convolution-less master equation in which time-varying damping rate functions characterize the correlation functions of the non-Markovian environment. To identify the damping rate function, we measure the time trace for one local observable of the spin chains which correspond to dynamical variables in a coherence vector representation. Although this formulation for the coherence vector is similar to that in~\cite{JunPRA15}, the dynamical equation for the coherence vector is time-varying due to the damping rate functions. Thus, the method in~\cite{JunPRA15} is invalid in our case.  We alternatively formulate the identification problem for the damping rate function as an optimization problem.  Also, we design a gradient algorithm to iteratively recover the damping rate functions for the non-Markovian environment.

Our contribution in this paper is that we provide a systematic approach to acquire  unknown damping rate functions in a time-convolution-less master equation for non-Markovian quantum systems. In principle, our gradient algorithm can achieve the damping rate function with a high fidelity, which would help to obtain an exact model of the spin chains in an experiment for quantum information processing. In addition, our method only requires to measure commutative observables of the spin chains, which is easy to be applied in an experiment. Because it avoids measuring non-commutative observables which cannot be accurately measured at the same time according to the uncertainty principle in quantum mechanics.

This paper is organized as follows. In Section \ref{II}, we introduce the time-convolution-less master equation for non-Markovian spin chains whose corresponding coherence vector representation is also introduced. The identification problem formulation and the gradient algorithm are given in Section \ref{III}. An example for two qubits in a non-Markovian environment is given in Section \ref{IV}. Conclusions are drawn in Section \ref{V}.
\section{Description of non-Markovian spin chains}\label{II}
\subsection{Time-convolution-less master equation}
\begin{figure}
  % Requires \usepackage{graphicx}
  \includegraphics[width=8.5cm]{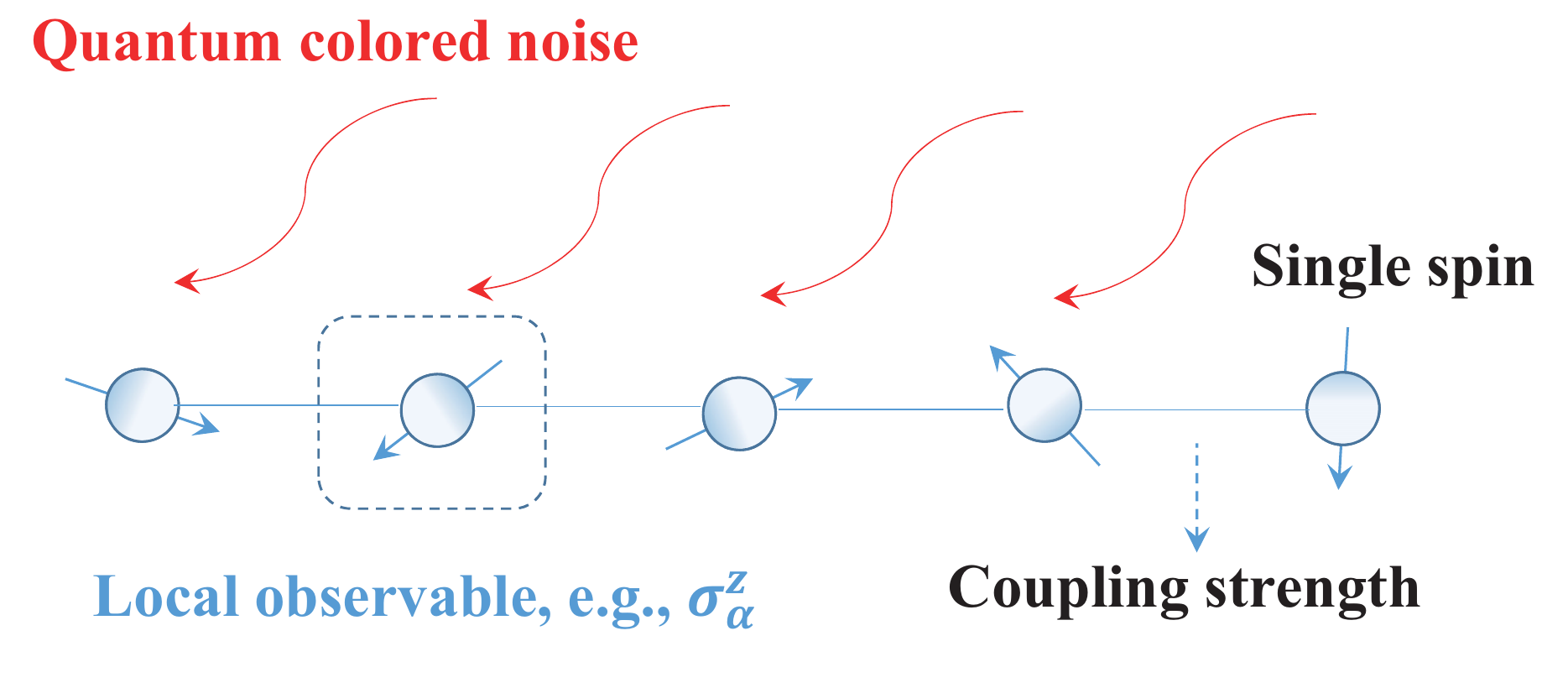}\\%figurefig.eps png
  \caption{Schematic plot of spin chains in a non-Markovian environment.}\label{SN}
\end{figure}
%A class of N-level quantum systems is a general quantum system model which can be used to describe one qubit or multi-qubit systems such as spin chains. When the N-level quantum systems are open to an environment with memory effects, the dynamics of the system are non-Markovian.
The non-Markovian spin chain we consider in this paper is plotted in Fig.\ref{SN} where the nodes and the edges represent spin components and their couplings, respectively. The components are disturbed by quantum colored noise arising from an unknown non-Markovian environment. Here, the time traces of local observables are measured and the resulting data is to be processed for identification.

In this paper, the non-Markovian dynamics of spin chains are described by a time-convolution-less master equation~\cite{Breuer}
\begin{equation}\label{1}
  \dot \rho(t)=\mathcal{L}_0\rho(t)+\mathcal{L}_{\gamma(t)}\rho(t),
\end{equation}
%where the density matrix $\rho(t)\in\mathbb{C}^{N\times N}$ describes the state
%with von Neumann properties $\rho(t)=\rho^\dagger(t)$, $\rho(t)\geq 0$, and ${\rm tr}[\rho(t)]=1$ represents the state
of the spin chains. %Here, $\rho^\dagger(t)$ is the Hermitian adjoint of $\rho(t)$.
The superoperator
\begin{equation}\label{2}
 \mathcal{L}_0\rho(t)=-i[H,\rho(t)]
\end{equation}
represents the internal dynamics of the spin chains where the Hamiltonian of the spin chains $H$ describes the internal energy of the system and their couplings. The commutator $[\cdot,\cdot]$ is calculated as $[A,B]=AB-BA$ for two matrices $A$ and $B$ with suitable dimensions. We let $\hbar=1$ hereafter.

The Lindblad dissipative term induced by the non-Markovian environment is expressed as
\begin{equation}\label{3}
 \mathcal{L}_{\gamma(t)}\rho(t)= \frac{1}{2}\sum_{j,k=1}^{N^2-1}\gamma_{jk}(t)([L_j,\rho(t) L_k^\dagger]+[L_j\rho(t), L_k^\dagger]).
\end{equation}
In contrast to the constant damping rates in Markovian master equations, the damping rate functions $\gamma_{jk}(t)$ for characterizing the correlation of the non-Markovian environment are  time-varying, which encapsule the coupling strengthes between the system and the environment and the density state of the environment~\cite{Breuer}. Here, we assume that $\gamma_{jk}(t)$ are complex functions, i.e., $\gamma_{jk}(t)={\rm Re}(\gamma_{jk}(t))+i{\rm Im}(\gamma_{jk}(t))$, where ${\rm Re}(\cdot)$ and ${\rm Im}(\cdot)$ represent real and imaginary parts of a function, respectively.

In addition, the coupling operators $L_k$ belong to a set $\mathcal{M}=\{L_k, k=1,\cdots, M=N^2-1\}$, which are an orthonormal basis for the Lie algebra $\mathfrak{su}(N)$.
%Also, $L_k$ is an $N\times N$ Hermitian matrix with $L_k=L_k^\dagger$ and ${\rm tr}(L_k)=0$. The Hilbert-Schmidt inner product is defined as $\langle L_m,L_n\rangle\equiv {\rm tr}(L^\dagger_mL_n)=\delta_{mn}, 1\leq m, n\leq M$.
Their commutation and anti-commutation relations can be calculated as
\begin{eqnarray}
% \nonumber to remove numbering (before each equation)
  [L_j,L_k] &=& i\sum_{l=1}^{M}C_{jkl}L_l, \label{4-1}\\
  \{L_j,L_k\} &=&\frac{2}{N}\delta_{jk}I+\sum_{l=1}^{M}D_{jkl}L_l,\label{4-2}
\end{eqnarray}
respectively. The coefficients  $C_{jkl},D_{jkl}\in\mathbb{R}$ are the completely antisymmetric and symmetric structure constants of the Lie algebra $\mathfrak{su}(N)$, i.e., with respect to interchange of any pair of indices~\cite{AL}. Here, the anti-commutator $\{\cdot,\cdot\}$ is calculated as $\{A,B\}=AB+BA$ for two matrices $A$ and $B$ with suitable dimensions and $\delta$ is the Kronecker delta function.

Note that although the variation of the density matrix in the time-convolution-less master equation (\ref{1}) only depends on the current density matrix mathematically, the time-varying damping rate functions enable (\ref{1}) to describe a non-Markovian behavior physically; i.e., the energy exchanges between the system and the environment~\cite{Breuer}.
%Also, with respect to different coupling operators, the system involves different dissipative channels such as dephasing channels. In addition, the damping rate functions fully carry the correlations of the environment and in the following we consider these damping rate functions to be unknown and are to be identified.

%Generally, the time-varying damping rate functions are derived from real parts of correlation functions of environment operators~\cite{} so we assume that $\gamma_{jk}(t)$ are real functions. The imaginary parts of correlation functions of environment operators can attribute Lamb shift terms which will not be considered in this paper. But the identification method for the damping rate functions can be applied to identified the Lamb shift terms.

\subsection{Dynamical equations in a coherence vector representation}
Under the orthonormal basis in $\mathcal{M}$, we can expand the density matrix $\rho(t)$ and the Hamiltonian $H$ as
\begin{eqnarray}
\rho(t)&=&\frac{1}{N}I_N+\sum_{n=1}^{M}x_n(t)L_n,\nonumber\\
H&=&\sum_{m=1}^M h_mL_m,\label{6}
\end{eqnarray}
respectively, where $h_m={\rm tr}(HL_m)\in\mathbb{R}$ and the coefficients $x_n(t)$ are the expectation values for the basis $L_n$; i.e., $x_n(t)={\rm tr}(\rho(t)L_n)\in\mathbb{R}$. They constitute the so-called coherence vector  $ \mathbf{x}(t)=[x_1(t),\cdots,x_n(t),\cdots,x_M(t)]^T\in\mathbb{R}^M$. Here, $I_N$ is an $N\times N$ identity matrix.

By using the relations (\ref{4-1}), (\ref{4-2}), and (\ref{6}), the time-convolution-less master equation (\ref{1}) can be rewritten as a dynamical equation in the coherence vector representation as
\begin{equation}\label{7}
 \dot {\mathbf{x}}(t)=\mathbf{A}(t)\mathbf{x}(t)+\mathbf{b}(t),~~x_n(0)={\rm tr}(L_n\rho(0)),
\end{equation}
where the initial value of the coherence vector $\mathbf{x}(0)$ is determined by the initial density matrix $\rho(0)$. The elements of the matrices $\mathbf{A}(t)\in\mathbb{R}^{M\times M}$ and $\mathbf{b}(t)\in\mathbb{R}^{M}$ can be calculated as
$A_{np}(t)=Q_{np}+R_{np}(t)$
with
% \nonumber to remove numbering (before each equation)
  $Q_{np} = \sum_{m=1}^MC_{mnp}h_m $,
 $R_{np}(t) = -\frac{1}{4}\sum_{l,j,k=1, l\leq m}^{M}(2-\delta_{jk}){\rm Re}(\gamma_{jk}(t))$$\times(C_{jln}C_{klp}+C_{kln}C_{jlp})$
$+\frac{1}{2}\sum_{l,j,k=1, l\leq m}^{M}{\rm Im}(\gamma_{jk}(t))$
$\times(C_{kln}D_{jlp}-C_{jln}D_{klp})$
and  $b_n(t)=-\frac{2}{N}\sum_{j,k=1, j<k}^{M}{\rm Im}(\gamma_{jk}(t))C_{jkn}$, respectively. Here, the corresponding coefficients $C$ and $D$ are introduced in the calculation of the commutation and anti-commutation relations (\ref{4-1}), (\ref{4-2}) when transforming Eq. (\ref{1}) to Eq. (\ref{7}). Their indexes correspond to the labels of the orthonormal basis for $\mathfrak{su}(N)$ involved in the commutation and anti-commutation relations.

Note that although Eq.~(\ref{7}) is in a similar form as the linear time-invariant dynamical equation for a Markovian quantum system~\cite{JunPRA15}, it is a linear time-varying dynamical equation due to the time-varying damping rate functions $\gamma_{jk}(t)$.
%The dynamics of the expectation value of an observable $L_n$, i.e., $x_n(t)=tr(L_n\rho(t))$ can be written as
%\begin{equation}\label{6}
%  \dot x_n(t)=\sum_{p=1}^{N^2-1}(Q_{np}+R_{np}(t))x_p(t)+b_n(t)
%\end{equation}
%where\begin{eqnarray}\label{7}
     % \nonumber to remove numbering (before each equation)
       %Q_{np} &=& \sum_{m=1}^M C_{mnp}h_m, \nonumber\\
    %   R_{np}(t) &=& -\frac{1}{4}\sum_{l,j,k=1}^{N^2-1}\gamma_{jk}(t)[C_{njl}(C_{klp}+D_{klp})+C_{knl}(C_{ljp}+D_{ljp})], \nonumber\\
     %  b_{n}(t)&=&\frac{1}{N}\sum_{j,k=1}^{N^2-1}{\rm Im}(\gamma_{jk}(t))C_{njk}.\nonumber
   %  \end{eqnarray}

On the other hand, to extract information of $\gamma_{jk}(t)$, one can measure expectation values of $S$ observables
\begin{equation}\label{9}
  \mathbf{y}(t)=[\langle O_1(t)\rangle, \langle O_2(t)\rangle,...,  \langle O_S(t)\rangle]^T,
\end{equation}
where $O_i(t)$ are local observables and the symbol $\langle\cdot\rangle={\rm tr}[\cdot\rho(t)]$ represents the expectation of an observable. Also, we can collect the data of $\mathbf{y}(t)$ as the output of the non-Markovian spin chains. In addition, these observables can be expanded under the orthonormal basis in $\mathcal{M}$ as $O_i=\sum_{n=1}^{M}o_{n}^{(i)}L_n$ with $o_{n}^{(i)}={\rm tr}(O_iL_n)$ and thus the output (\ref{9}) can be reexpressed as
\begin{equation}\label{10}
  \mathbf{y}(t)=\mathbf{c}\mathbf{x}(t)
\end{equation}
where $o_{n}^{(i)}$ is the $i$th row and $n$th column element of the matrix $\mathbf{c}\in\mathbb{R}^{S\times M}$.

Note that the measured operators in $\mathbf{y}(t)$ should be commutative since the uncertainty principle in quantum mechanics requires that two non-commutative operators cannot be measured accurately at the same time. Hence, in general, the number of the measured observables $S$ should be less than that of the orthonormal basis in $\mathcal{M}$, i.e., $S<M$.

It should be noted that the measurement of expectations of observables is applicable to non-Markovian quantum systems. For example, in a superconducting non-Markovian single qubit system, the measurement of average charge numbers is equivalent to measuring the $z$-component of the coherence vector for the single qubit. The procedure is to measure the same observable many times and then take the average of the results~\cite{PhysRevA.87.022324}.
\subsection{Reduced dynamical equation}
The equation (\ref{7}) describes the full dynamics of the non-Markovian spin chains. However, it is possible that not all the components in $\mathbf{x}(t)$ are relevant to the observables (\ref{9}), i.e., the matrix $\mathbf{A}(t)$ would have a block-diagonalizable structure and a proper sub-block corresponds to the observables. We can use a filtration procedure in~\cite{Sastry} to find the {\it accessible set} of the which is related to the observables~\cite{JunPRA15}.

%An orthonormal basis $\mathcal{M}'=\{L_{k'_1},\cdots,L_{k'_{M'}}\}$ can be defined, which expands the observables (\ref{9}). It can be a subset of $\mathcal{M}$, i.e., $\mathcal{M}'\subseteq\mathcal{M}$ with $M'\leq M$. To obtain all the components in the coherence vector corresponding to the observables (\ref{9}),  an iterative procedure~\cite{Sastry}
%\begin{equation}\label{11}
%  G_i=\mathcal{L}^\dagger[G_{i-1}]\cup G_{i-1},~~G_0=\mathcal{M}',
%\end{equation}
%can be applied, where $\mathcal{L}^\dagger[G_{i-1}]=\{L_j: {\rm tr}(L_j^\dagger\mathcal{L}^\dagger g)\neq 0~~{\rm where}~~g\in G_{i-1}\}$. The adjoint superoperator $\mathcal{L}^\dagger$ can be calculated as
%\begin{eqnarray}\label{12}
%% \nonumber to remove numbering (before each equation)
%  \mathcal{L}^\dagger X&=& -i[X,H] \nonumber\\
%   &&+\frac{1}{2}\sum_{j,k=1}^{N^2-1}\gamma_{jk}([L_j,L^\dagger_k X]+[XL_j, L_k^\dagger]),
%\end{eqnarray}
%for an operator $X$. The iterative procedure will stop when no new element is added to $G_i$. The final set $G_i$ contains the basis which is related to the observables and referred to as the {\it accessible set}~\cite{JunPRA15}.

Thus, a reduced coherence vector $\tilde{\mathbf{x}}(t)$ corresponding to the observables (\ref{9}) can be defined and it obeys a reduced dynamical equation
\begin{eqnarray}\label{13}
% \nonumber to remove numbering (before each equation)
\dot{\tilde {\mathbf{x}}}(t) &=& \tilde{ \mathbf{A}} (t){\tilde {\mathbf{x}}}(t)+{\tilde{ \mathbf{b}}}(t),\nonumber\\
 { \mathbf{y}}(t)&=&\tilde{\mathbf{c}}\tilde {\mathbf{x}}(t),
\end{eqnarray}
where $\tilde{\mathbf{A}} (t)$, $\tilde{ \mathbf{b}}(t)$, and $\tilde{ \mathbf{c}}$ with suitable dimensions are sub-matrices of $ \mathbf{A} (t)$, ${ \mathbf{b}}(t)$, and ${ \mathbf{c}}$.

%Note that the reduced dynamical equation provides low-dimensional dynamics with respect to the observables, which will aid the identification process.
\section{Gradient algorithm for identifying the damping rate function}\label{III}
\subsection{Optimization formulation for the identification of the damping rate function}

The identification problem considered in this paper can be stated as follows. Considering the non-Markovian spin chains governed by the dynamical equation of the coherence vector (\ref{7}), where all the information of the environment are encoded in the unknown damping rate functions, the environment identification problem is to identify the time-varying damping rate functions from the data of the time traces of the observables $\hat {\mathbf{y}}(t)$ (\ref{10}).

Note that due to the interactions between the spin chains and the environment, in general, there is no decoherence-free subspace for a common non-Markovian system without control pulses; i.e., decoherence channels affect all the components in the coherence vector. Thus, the damping rate functions will be imprinted in the observables. Hence, the identification problem for the damping rate function should be solvable. In addition, since the time-varying damping rate functions $\gamma_{jk}(t)$ result in a linear time-varying system (\ref{7}), the identification method for Markovian quantum systems in~\cite{JunPRA15} cannot be applied in our case.

Generally, it is difficult to identify analytically a time-varying function in a dynamical system. Alternatively, we can numerically solve this problem by designing an algorithm. We consider that the system (\ref{7}) evolves in a total time $T$ during which the observables are measured $K$ times at equal time intervals $\Delta t$, i.e., $\Delta t=\frac{T}{K}$. Thus, a set of real measurement results $\hat{\mathbf{y}}=[\hat{\mathbf{y}}(0),\hat{\mathbf{y}}(1),\cdots,\hat{\mathbf{y}}(K-1)]$ can be obtained.

Also, the time-varying damping rate functions $\gamma_{jk}(t)$ are discretized and we assume the $\gamma_{jk}(t)$ in each time interval are  constants $\gamma_{jk}(\kappa), \kappa=0,1,2,\cdots,(K-2)$. Thus in each time interval, we can consider the linear time varying (LTV) system equation (\ref{7}) as a linear time-invariant (LTI) system and write the dynamical equation in a continuous-time form as
\begin{eqnarray}
 \dot{ \mathbf{x}}(\tau)&=&  \mathbf{A}(\gamma_{jk}(\tau))\mathbf{x}(\tau)+\mathbf{b}({\rm Im}(\gamma_{jk}(\tau))), \label{14-1}\\
  \mathbf{y}(\tau)&=&\mathbf{c}\mathbf{x}(\tau),~~~~~~~~~\tau\in[\kappa\Delta t,(\kappa+1)\Delta t] \label{14-2}
\end{eqnarray}
where Eq.~(\ref{14-1}) can be solved as
\begin{eqnarray}\label{15}
  &&\mathbf{x}(\kappa+1)=e^{\mathbf{A}(\gamma_{jk}(\kappa))\Delta t}\mathbf{x}(\kappa)\nonumber\\
  &&+\int_{\kappa\Delta t}^{(\kappa+1)\Delta t}e^{\mathbf{A}(\gamma_{jk}(\kappa))((\kappa+1)\Delta t-\tau)}\mathbf{b}({\rm Im}(\gamma_{jk}(\kappa))){\rm d}\tau.\nonumber\\ %e^{\mathbf{A}(\gamma_{jk}(\kappa))((\kappa+1)\Delta t-\tau)}
\end{eqnarray}
Here, we have written the initial coherence vector in the current and next time intervals in an abbreviate form as $\mathbf{x}(\kappa)$ and $\mathbf{x}(\kappa+1)$, respectively.

For a given set of $\gamma_{jk}(\kappa), \kappa=0,1,2,\cdots,(K-2)$, an output  vector $\mathbf{y}=[\mathbf{y}(0),\mathbf{y}(1),\cdots,\mathbf{y}(K-1)]$ can be generated by using (\ref{14-2}) and (\ref{15}).
A guessed set of $\gamma_{jk}(\kappa), \kappa=0,1,2,\cdots,(K-2)$ may not be identical to the real damping rate function. Hence, the generated output $\mathbf{y}$ will be different from the real data $\hat{\mathbf{y}}$. So we define an objective function
\begin{equation}\label{16}
 J=\frac{1}{2}\sum_{\kappa=0}^{K-1}(\mathbf{y}(\kappa)-\hat{ \mathbf{y}}(\kappa))^2
\end{equation}
to evaluate the distance between the two vectors $\mathbf{y}$ and $\hat{\mathbf{y}}$.

Therefore, the identification problem for the damping rate functions $\gamma_{jk}(t)$ can be converted into an optimization problem as follows.

For real measurement results $\hat{\mathbf{y}}$, the optimization problem is to find a set of $\gamma_{jk}(\kappa), \kappa=0,1,2,\cdots,(K-2)$ such that the objective function (\ref{16}) is minimized, that is
\begin{eqnarray}\label{17}
&\min\limits_{\gamma_{jk}(\kappa)}J&\nonumber\\
{\rm s.~t.}&{\rm (\ref{7}), (\ref{10})}.&
\end{eqnarray}
\subsection{Gradient algorithm for the identification problem}
To design a gradient algorithm for the optimization problem (\ref{17}),
it is crucial to obtain the gradient of the objective $J$ with
respect to $\gamma_{jk}(\kappa)$ in each time interval. By using the chain rule,
the gradient can be calculated as
\begin{eqnarray}\label{18}
\frac{{\rm d}J}{{\rm d}\gamma_{jk}(\kappa)}&=&\frac{{\rm d}J}{{\rm d}\mathbf{y}}\cdot\frac{{\rm d}\mathbf{y}}{{\rm d}\mathbf{x}}\cdot\frac{{\rm d}\mathbf{x}}{{\rm d}\gamma_{jk}(\kappa)}\nonumber\\
&=&\sum_{\kappa'=0}^{K-1}(\mathbf{y}(\kappa')-\hat{ \mathbf{y}}(\kappa'))\cdot\mathbf{c}\frac{{\rm d}\mathbf{x}(\kappa')}{{\rm d}\gamma_{jk}(\kappa)}
\end{eqnarray}
where
\begin{equation}\label{19}
  \frac{{\rm d} \mathbf{x}(\kappa')}{{\rm d} \gamma_{jk}(\kappa)}=\left\{\begin{array}{ll}
0&\kappa'<\kappa+1\\
 \frac{{\rm d} \mathbf{x}(\kappa+1)}{{\rm d} \gamma_{jk}(\kappa)}& \kappa'=\kappa+1.\\
\prod_{\kappa''=\kappa+1}^{\kappa'-1} e^{\mathbf{A}(\gamma_{jk}(\kappa''))\Delta t}  \frac{{\rm d} \mathbf{x}(\kappa+1)}{{\rm d} \gamma_{jk}(\kappa)}&
  \kappa'>\kappa+1
\end{array}\right.
\end{equation}

In the following, we will calculate the derivative of $\mathbf{x}$ with respect to the real and imaginary parts of $\gamma_{jk}(\kappa)$, respectively. By using a standard formula for
derivative of a matrix exponential~\cite{sfexp}
\begin{equation}\label{20}
  \frac{{\rm d}}{{\rm d}x}e^{(A+xB)t}|_{x=0}=e^{At}\int_0^t e^{A\tau}Be^{-A\tau}{\rm d}\tau
\end{equation}%
%\begin{equation}\label{20}
%\frac{{\rm d}}{{\rm d}h}e^{t(A+hV)}|_{h=0}=e^{tA/2}\int_{-t/2}^{t/2}e^{-\tau A}Ve^{\tau A}{\rm d}\tau e^{tA/2}
%\end{equation}
with matrices $A$ and $B$ of suitable dimensions,
%\begin{equation}\label{20}
%  \frac{\rm d}{{\rm d}x}e^{(A+xB)t}|_{x=0}=e^{At}\int_0^te^{A\tau}Be^{-A\tau}{\rm d}\tau,
%\end{equation}
we  have
\begin{eqnarray}\label{21}
  \frac{{\rm d}e^{\mathbf{A}(\gamma_{jk}(\kappa))\Delta t}}{{\rm d}{\rm Re}(\gamma_{jk}(\kappa))}&=&\Delta te^{\mathbf{A}(\gamma_{jk}(\kappa))\Delta t}\mathbf{\tilde E_R},\nonumber\\
  \frac{{\rm d}e^{\mathbf{A}(\gamma_{jk}(\kappa))\Delta t}}{{\rm d}{\rm Im}(\gamma_{jk}(\kappa))}&=&\Delta te^{\mathbf{A}(\gamma_{jk}(\kappa))\Delta t}\mathbf{\tilde E_I},
\end{eqnarray}
where $\mathbf{\tilde E_R}=\int_{\kappa\Delta t}^{(\kappa+1)\Delta t}e^{\mathbf{A}(\gamma_{jk}(\kappa))\tau}\mathbf{E_R}e^{-\mathbf{A}(\gamma_{jk}(\kappa))\tau}{\rm d}\tau$ and $\mathbf{\tilde E_I}=\int_{\kappa\Delta t}^{(\kappa+1)\Delta t}e^{\mathbf{A}(\gamma_{jk}(\kappa))\tau}\mathbf{E_I}e^{-\mathbf{A}(\gamma_{jk}(\kappa))\tau}{\rm d}\tau$. When the time interval $\Delta t$ is small enough such that $\Delta t\ll ||\mathbf{A}||^{-1}$, we have
\begin{equation}\label{22}
  \mathbf{\tilde E_R}=\mathbf{E_R},~~\mathbf{\tilde E_I}=\mathbf{E_I},
\end{equation}
where the elements of $\mathbf{E_R}$  and $\mathbf{E_I}$ can be expressed as $E_{Rnp}=-\frac{1}{4}(2-\delta_{jk})\sum_{l=1}^{M}(C_{jln}C_{klp}+C_{kln}C_{jlp})$ and  $E_{Inp}=\frac{1}{2}\sum_{l=1}^M(C_{kln}D_{jlp}-C_{jln}D_{klp})$.

Thus, the derivative of $\mathbf{x}(\kappa+1)$ with respect to the real and imaginary parts of $\gamma_{jk}(\kappa)$ can be calculated as
\begin{eqnarray}\label{23}
&&\frac{{\rm d} \mathbf{x}(\kappa+1)}{{\rm d} {\rm Re}(\gamma_{jk}(\kappa))}=\Delta te^{\mathbf{A}(\gamma_{jk}(\kappa))\Delta t}\mathbf{E_R}\mathbf{x}(\kappa)\nonumber\\
&&+\int_{\kappa\Delta t}^{(\kappa+1)\Delta t}((\kappa+1)\Delta t-\tau)e^{\mathbf{A}(\gamma_{jk}(\kappa))((\kappa+1)\Delta t-\tau)}\nonumber\\
&&\times\mathbf{E_R}\mathbf{b}({\rm Im}\gamma_{jk}(\kappa)){\rm d}\tau,
\end{eqnarray}
\begin{eqnarray}\label{24}
&&\frac{{\rm d} \mathbf{x}(\kappa+1)}{{\rm d} {\rm Im}(\gamma_{jk}(\kappa))}=\Delta te^{\mathbf{A}(\gamma_{jk}(\kappa))\Delta t}\mathbf{E_I}\mathbf{x}(\kappa)\nonumber\\
&&+\int_{\kappa\Delta t}^{(\kappa+1)\Delta t}((\kappa+1)\Delta t-\tau)e^{\mathbf{A}(\gamma_{jk}(\kappa))((\kappa+1)\Delta t-\tau)}\nonumber\\
&&\times\mathbf{E_I}\mathbf{b}({\rm Im}\gamma_{jk}(\kappa)){\rm d}\tau\nonumber\\
&&+\int_{\kappa\Delta t}^{(\kappa+1)\Delta t}e^{\mathbf{A}(\gamma_{jk}(\kappa))((\kappa+1)\Delta t-\tau)}\mathbf{F}{\rm d}\tau,
\end{eqnarray}
where the $n$th row element of the column vector $\mathbf{F}$ is $-\frac{2}{N}C_{jkn}$.
%and
%
%\begin{equation}\label{21}
%  \delta\Phi(\gamma_{jk}(\kappa),\Delta t)=\delta\gamma_{jk}(\kappa)\Delta t\Phi(\Delta t)\int_{0}^{\Delta t}\Phi(-\tau)\mathbf{E}\Phi(\tau)d\tau
%\end{equation}
%which can be simplified as
%\begin{equation}\label{21}
%  \delta\Phi(\gamma_{jk}(\kappa),\Delta t)=\delta\gamma_{jk}(\kappa)\Delta t\Phi(\Delta t)\mathbf{E}
%\end{equation}
%when $\Delta t\ll ||\mathbf{A}||^{-1}$

By combining (\ref{18}), (\ref{19}), (\ref{23}), and (\ref{24}), we obtain the gradient of the objective $J$ with respect to $\gamma_{jk}$. Therefore, if we update $\gamma_{jk}$ as
\begin{eqnarray}\label{25}
  {\rm Re}(\gamma_{jk})\rightarrow {\rm Re}(\gamma_{jk})-\epsilon_R\cdot\frac{{\rm d}J}{{\rm d} {\rm Re}(\gamma_{jk})},\nonumber\\
    {\rm Im}(\gamma_{jk})\rightarrow {\rm Im}(\gamma_{jk})-\epsilon_I\cdot\frac{{\rm d}J}{{\rm d} {\rm Im}(\gamma_{jk})},
\end{eqnarray}
with small step sizes $\epsilon_R$ and $\epsilon_I$, we can minimize the objective $J$. With the relation (\ref{25}), our gradient algorithm for the identification problem can be summarized as follows.

{\it Step 1}. Choose and measure a set of observables $\hat{\mathbf{y}}(t)$; Initialize the state of the system $\mathbf{x}(0)$, the output $\mathbf{y}(0)$, and the step sizes $\epsilon_R$ and $\epsilon_I$; Guess the initial values of $\{\gamma_{jk}(\kappa)\}$;

{\it Step 2}. Calculate $\mathbf{x}(1)$ to $\mathbf{x}(K-1)$ and $\mathbf{y}(1)$ to $\mathbf{y}(K-1)$ with their initial values ${\mathbf{x}}(0)$ and $\mathbf{y}(0)$;

{\it Step 3}. Compute the gradient of the objective $J$ with respect to $\gamma_{jk}(\kappa)$;

{\it Step 4}. Update  $\gamma_{jk}(\kappa)$ by using the relation (\ref{25});

{\it Step 5}. When a termination condition is satisfied, stop the algorithm; Otherwise, go to the {\it Step 2} and start a new iteration.

Note that this algorithm searches optimal solutions according to the gradient and hence it may  stop at a local minimum point. Moreover, the final performance of the algorithm depends on the initial guessed solution $\{\gamma_{jk}(\kappa)\}$. When the initial guessed solution is closer to the optimal one, the algorithm will converge faster. The step size may also affect the performance of the algorithm. A small step size would result in a slow convergence process and a large one would lead to an algorithm that saturates in the  vicinity of a minimum. An adaptive step size can avoid the above problems. In addition, a proper termination condition can be the completion of a given number of iterations or the attainment of an accuracy of the objective $J$.

\section{A Physical Example}\label{IV}
In the example, we consider two coupled qubits immersed in an unknown common environment where the dissipative processes for the two qubits can be considered to be the same. Its non-Markovian dynamics can be described by a time-convolution-less master equation~\cite{Breuer} as
\begin{eqnarray}\label{50}
\dot{\rho}_q(t)&=&-i[H_q,\rho_q(t)]+\nonumber\\
&&\frac{\gamma(t)}{2}\sum_{l=1}^2([\sigma^-_l\rho_q(t),\sigma^+_l]+[\sigma^-_l,\rho_q(t)\sigma^+_l]),
% [2\sigma^-_\alpha\rho_q(t)\sigma^+_\alpha-\sigma^+_\alpha\sigma^-_\alpha\rho_q(t)-\rho_q(t)\sigma^+_\alpha\sigma^-_\alpha]
  %&&+\frac{\Delta(t)-\mu(t)}{2}\sum_{\alpha=1}^2([\sigma^+_\alpha\rho_q(t),\sigma^-_\alpha]+[\sigma^+_\alpha,\rho_q(t)\sigma^-_\alpha]).\nonumber\\
  %[2\sigma^+_\alpha\rho_q(t)\sigma^-_\alpha-\sigma^-_\alpha\sigma^+_\alpha\rho_q(t)-\rho_q(t)\sigma^-_\alpha\sigma^+_\alpha]
\end{eqnarray}
where $\rho_q(t)$ is the density operator for the two qubits and the ladder operators are defined as $\sigma^+=\frac{1}{2}(\sigma^x+i\sigma^y)$ and $\sigma^-=\frac{1}{2}(\sigma^x-i\sigma^y)$ with Pauli matrices
\begin{equation}\label{28}
\sigma^x= \left[ \begin{array}{cc}
0 &1 \\
1 & 0 \end{array} \right],~~
\sigma^y= \left[ \begin{array}{cc}
0 &-i \\
i & 0\end{array} \right],~~
\sigma^z=\left[ \begin{array}{cc}
1 &0\\
0 & -1 \end{array} \right].
\end{equation}
Here, the label $l$ is used to index the qubits. The two qubits are assumed to be coupled in an XY interaction form and thus the Hamiltonian of the two qubits is written as
\begin{equation}\label{51}
 H_q=\sum_{\alpha=1}^2\frac{\omega_\alpha}{2}\sigma^z_\alpha+\frac{g}{2}(\sigma^x_1\sigma^x_2+\sigma^y_1\sigma^y_2),
\end{equation}
where $g$ is the coupling strength between two qubits and $\omega_\alpha$ is the angular frequency for the $\alpha$-th qubit.
In this time-convolution-less master equation (\ref{50}), all the properties of the unknown environment are combined in the damping rate function $\gamma(t)$ which are the same with respect to each dissipative channel for the two qubits. Thus the identification task is to recover the unknown damping rate function $\gamma(t)$. To simulate a real damping rate function induced by the non-Markovian environment, we assume that the real damping rate function is
\begin{equation}\label{29}
\hat \gamma(t)=\frac{2h\lambda{\rm sinh}(dt/2)}{d{\rm cosh}(dt/2)+\lambda{\rm sinh}(dt/2)},
\end{equation}
which results from a non-Markovian environment with a Lorentzian spectrum~\cite{Breuer}. The values of the parameters $h$, $\lambda$, and $d$ will be given in the following paragraph. This real damping rate function (\ref{29}) is utilized to generate the real measurement results $\hat{\mathbf{y}}(t)$.

Furthermore, we measure the observable $\sigma^z_1$ for the first qubit which induces the accessible set $\{\sigma^z_1,\sigma^z_2,\sigma^x_1\sigma^x_2,\sigma^x_1\sigma^y_2,\sigma^y_1\sigma^x_2,\sigma^y_1\sigma^y_2\}$. Due to the interactions between the two qubits, the accessible set also includes the operators for the second qubit. Hence, the reduced dynamical equation for the state vector $\tilde{\mathbf{x}}(t)=[\langle\sigma^z_1\rangle,\langle\sigma^z_2\rangle,\langle\sigma^x_1\sigma^x_2\rangle,\langle\sigma^x_1\sigma^y_2\rangle,\langle\sigma^y_1\sigma^x_2\rangle,\langle\sigma^y_1\sigma^y_2\rangle]^T$ is in the form of (\ref{13}) with matrices
\begin{eqnarray}\label{52}
% \nonumber to remove numbering (before each equation)
  &&\tilde{\mathbf{A}}(t) =\nonumber\\
  &&\left[
                            \begin{array}{cccccc}
                              -\gamma(t) & 0 & 0 & -g & g & 0 \\
                              0 & -\gamma(t)& 0 & g & -g & 0 \\
                              0 & 0 & -\gamma(t) & -\omega_2 & -\omega_1 & 0\\
                               g & -g & \omega_2& -\gamma(t) & 0 & -\omega_1 \\
                              -g & g & \omega_1 & 0 & -\gamma(t) & -\omega_2\\
                              0 & 0 & 0 & \omega_1 &\omega_2 & -\gamma(t) \\
                            \end{array}
                          \right],\nonumber
\end{eqnarray}
\begin{equation}\label{52}
  \tilde{\mathbf{b}}(t) = \left[
                            \begin{array}{cccccc}
                              -\gamma(t) &  -\gamma(t)  & 0 & 0 & 0 & 0 \\
                            \end{array}
                          \right]^T.\nonumber
\end{equation}

 \begin{figure}
  % Requires \usepackage{graphicx}
  \includegraphics[width=8.5cm]{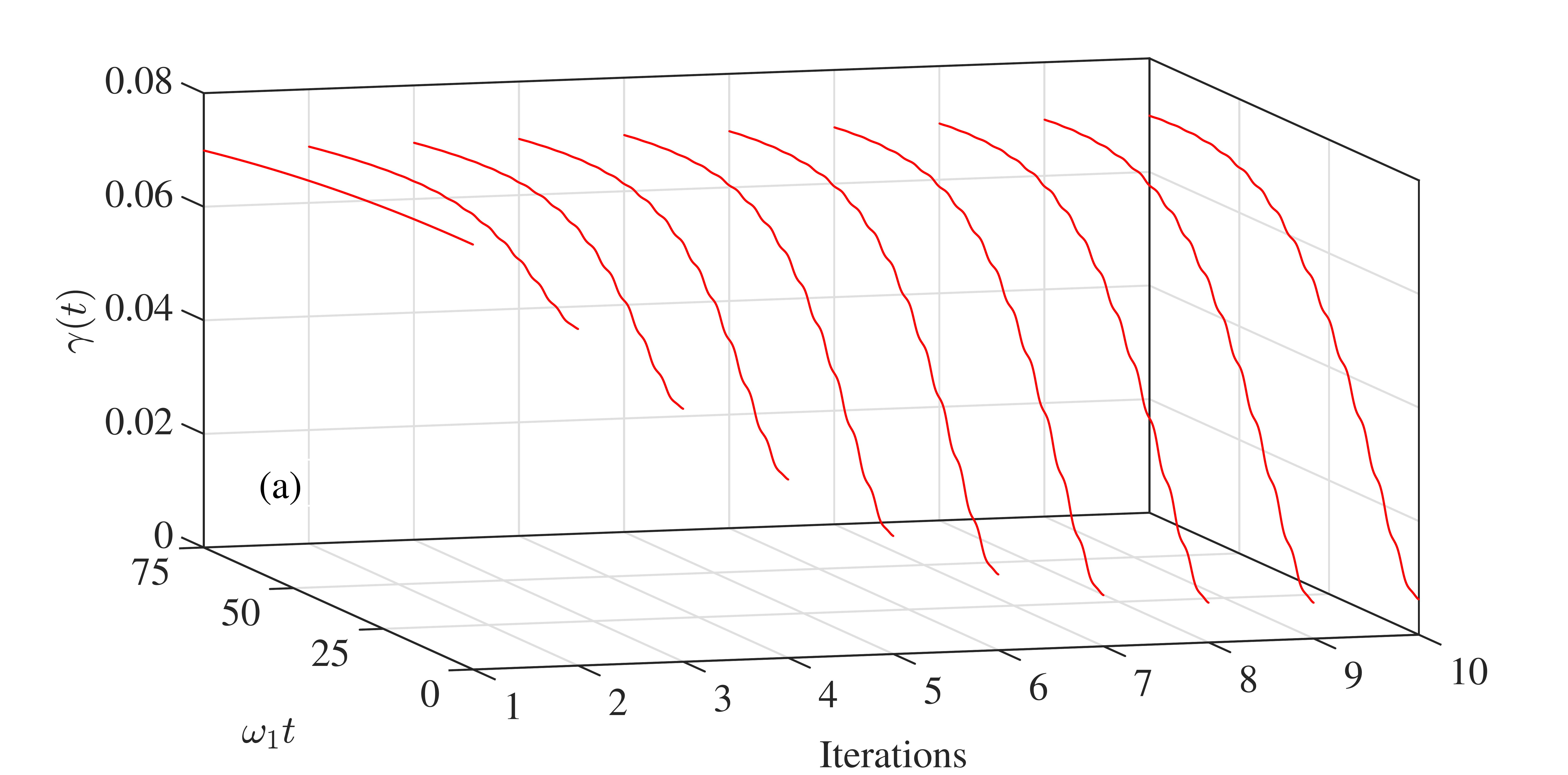}\\%figurefig.eps png
    \includegraphics[width=8.5cm]{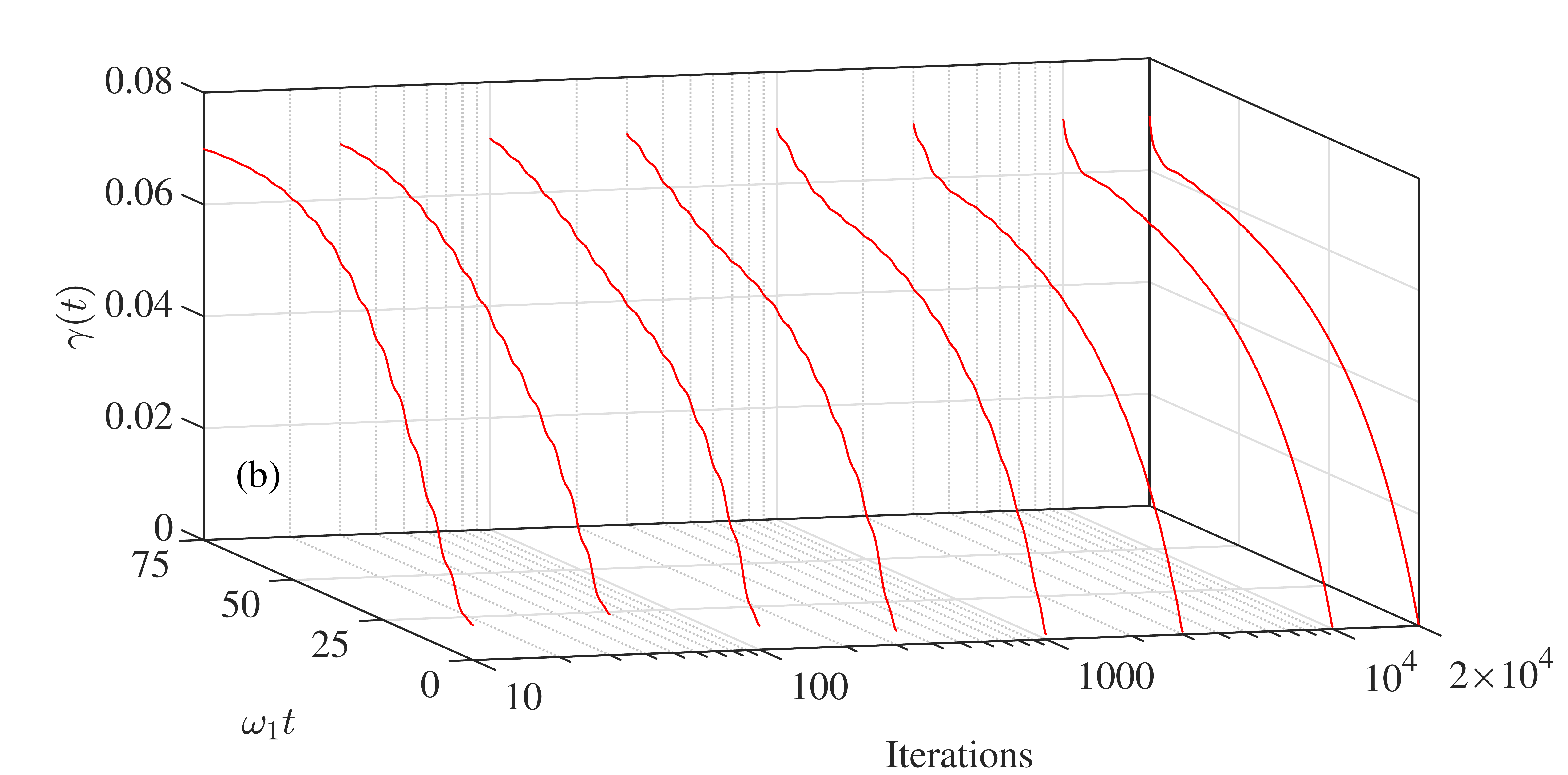}
  \caption{Evolution of the identified damping rate function at specific iterations.}\label{f2}
\end{figure}

In addition, the parameters of the system are chosen as follows. The angular frequencies of the two qubits are the same; i.e., $\omega_1=\omega_2=1.5{\rm GHz}$. The coupling strength between the two qubits is $g=1{\rm GHz}$. To simulate the real $\gamma(t)$, the parameters in (\ref{29}) are chosen as $h = 0.05{\rm GHz}$, $d = 0.05{\rm GHz}$ and $\lambda=0.1{\rm GHz}$. The gradient algorithm starts with a guessed damping rate function
\begin{equation}\label{51-1}
  \gamma_{0}(t)=0.04\cos(0.01t)+0.0348,
\end{equation}
which is plotted as a red line at the first iteration in Fig. \ref{f2}(a).

Both the step size in each iteration and the termination condition can affect the performance of the gradient algorithm. We choose the step size for updating the damping rate function as $\epsilon=0.002$. Because a large step size would result in oscillations in the final stage for searching the minimum $J$ and a small one would lead to a slow convergence process. On the other hand, we choose to iterate the algorithm for 20000 times as the terminal condition. Sufficient iteration times allow the algorithm to obtain a minimal objective $J$.

\begin{figure}
  % Requires \usepackage{graphicx}
  \includegraphics[width=8.5cm]{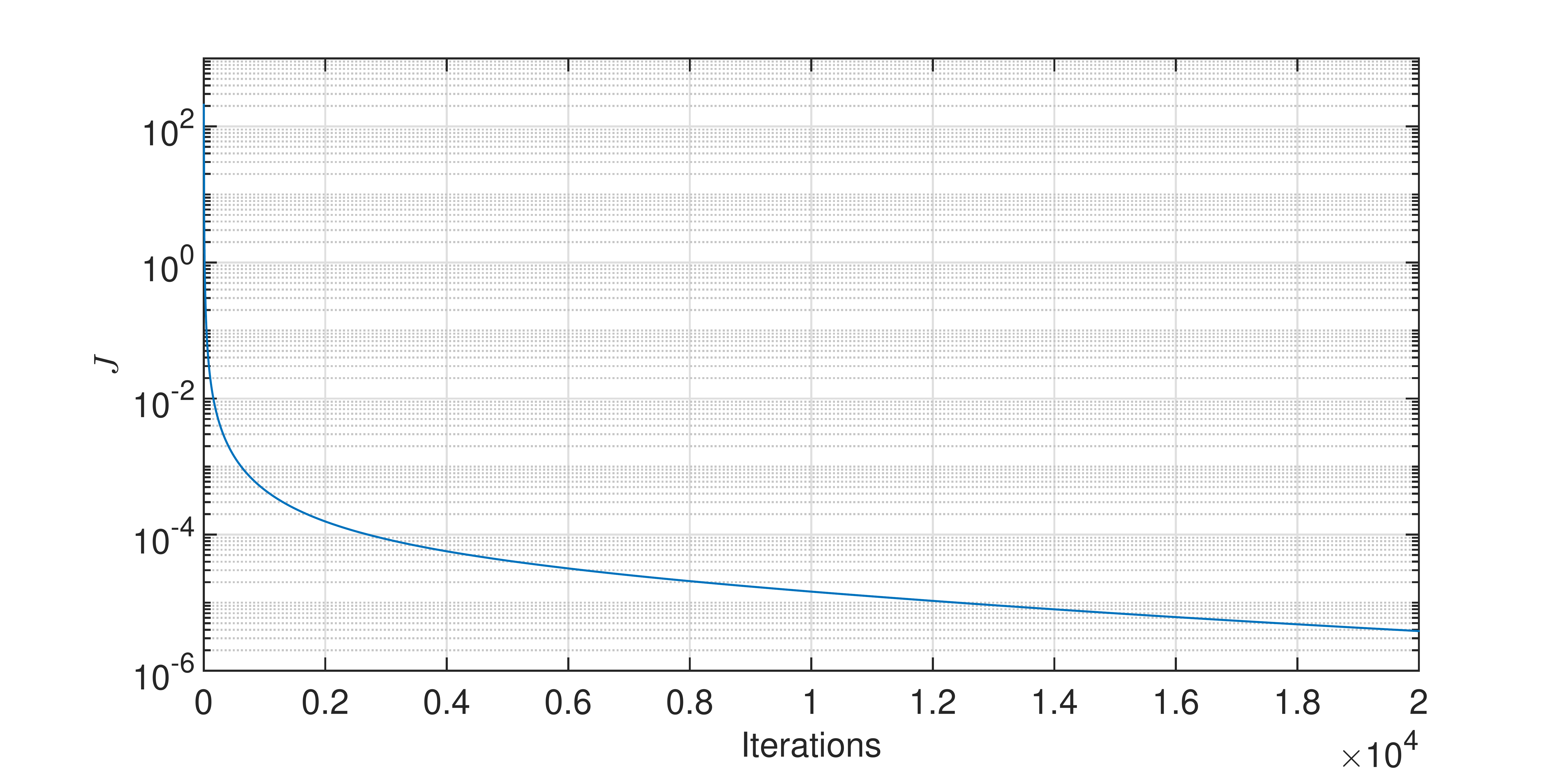}\\%figurefig.eps png
  \caption{Reduction of the objective function $J$.}\label{f3}
\end{figure}

With these parameters, the evolution of the damping rate function from the initial guess to the identified one are given in Fig. \ref{f2}. Initially, the gradient algorithm can significantly update the identified damping rate function as shown in Fig. \ref{f2}(a). After $10$ iterations, the convergence rate of the identification process slows down as shown in Fig. \ref{f2}(b). This phenomenon also reflects in the reduction of the objective function $J$ in Fig. \ref{f3}.
With 20000 iterations, the objective function is down to about $10^{-5}$ and we terminate the algorithm.

\begin{figure}
  % Requires \usepackage{graphicx}
  \includegraphics[width=8.5cm]{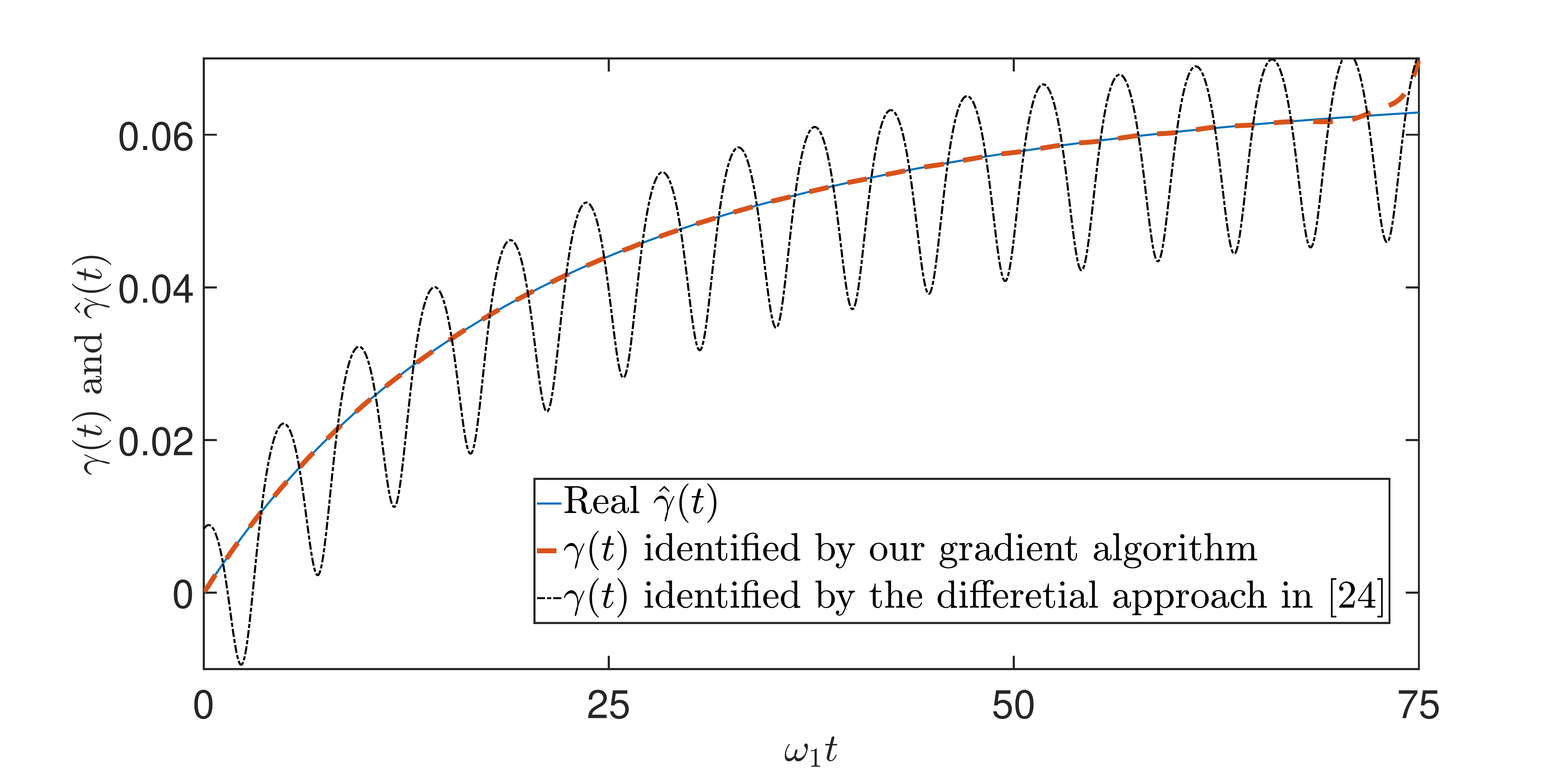}\\%figurefig.eps png
  \caption{Comparison between the identified and real damping rate functions.}\label{f4}
\end{figure}

The final identified damping rate function is plotted as the dashed red line in Fig. \ref{f4}. Compared to the real damping rate function represented as the solid blue line, our algorithm can identify the damping rate function well except at the final stage. This is because the measurement results at a final time interval contain limited information of the damping rate function and thus the resulting gradient provides very small updates. We also make a comparison between our method and a differential approach in~\cite{PhysRevA.82.062104} whose identified result is plotted as the dashed dot line in Fig.~\ref{f4}. Under the identical measurement result, our method obtains a better result than that by using the differential approach. This is because all the corresponding measured results are utilized to identify the unknown damping rate function in a sample time interval by using our method. However, the differential approach only use the measurement results at the current and next times to estimate the damping rate function at the current time. Its identification result can be improved if we measure the observable densely.

%\begin{figure*}[!t]
%\begin{eqnarray}\label{52}
%% \nonumber to remove numbering (before each equation)
%  \tilde{\mathbf{A}}(t) &=& \left[
%                            \begin{array}{cccccc}
%                              -\gamma(t) & 0 & 0 & -g & g & 0 \\
%                              0 & -\gamma(t)& 0 & g & -g & 0 \\
%                              0 & 0 & -\gamma(t) & -\omega_2 & -\omega_1 & 0\\
%                               g & -g & \omega_2& -\gamma(t) & 0 & -\omega_1 \\
%                              -g & g & \omega_1 & 0 & -\gamma(t) & -\omega_2\\
%                              0 & 0 & 0 & \omega_1 &\omega_2 & -\gamma(t) \\
%                            \end{array}
%                          \right]
%  \nonumber \\
%  \tilde{\mathbf{b}}(t) &=& \left[
%                            \begin{array}{cccccc}
%                              -\gamma(t) &  -\gamma(t)  & 0 & 0 & 0 & 0 \\
%                            \end{array}
%                          \right]^T.
%\end{eqnarray}
%\end{figure*}

\section{Conclusions}\label{V}
In this paper, we have designed a gradient algorithm for
identifying the damping rate function in the time-convolution-less master equation for non-Markovian spin chains. We have formulated the identification problem as an optimization problem that can be solved iteratively
by calculating the gradient of the objective with respect to the
damping rate function in each time interval. The numerical example on a non-Markovian two-qubit system demonstrates the efficacy of our gradient algorithm. Compared to a differential approach, our method can identify the damping rate function with a high fidelity.

\end{document}